\documentclass[conference]{IEEEtran}
\IEEEoverridecommandlockouts
\usepackage{cite}
\usepackage[colorlinks=true, citecolor=black, urlcolor=black, linkcolor=black]{hyperref}

\usepackage{amsmath,amssymb,amsfonts}
\usepackage{algorithmic}
\usepackage{graphicx}
\usepackage{textcomp}
\usepackage{xcolor}
\usepackage{lipsum}
\def\BibTeX{{\rm B\kern-.05em{\sc i\kern-.025em b}\kern-.08em
    T\kern-.1667em\lower.7ex\hbox{E}\kern-.125emX}}
\begin{document}

\title{Lattice XBAR Filters in Thin-Film Lithium Niobate \\
\thanks{This work was supported by the Defense Advanced Research Projects Agency (DARPA) under the Compact Front-End Filters at the Element-Level (COFFEE) program No.\ HR0011-22-2-0031, the National Science Foundation (NSF) under CAREER Award No.\ 2339731, and the Anandamahidol Foundation Scholarship.}
}

\author{Taran Anusorn, Byeongjin Kim, Ian Anderson, Ziqian Yao, and Ruochen Lu \\
Chandra Family Department of Electrical and Computer Engineering \\
The University of Texas at Austin, Austin, TX, USA \\
{taran.anusorn@utexas.edu} }

\maketitle

\begin{abstract}
This work presents the demonstration of lattice filters based on laterally excited bulk acoustic resonators (XBARs). Two filter implementations, namely direct lattice and layout-balanced lattice topologies, are designed and fabricated in periodically poled piezoelectric film (P3F) thin-film lithium niobate (TFLN). By leveraging the strong electromechanical coupling of XBARs in P3F TFLN together with the inherently wideband nature of the lattice topology, 3-dB fractional bandwidths (FBWs) of 27.42\% and 39.11\% and low insertion losses (ILs) of 0.88 dB and 0.96 dB are achieved at approximately 20 GHz for the direct and layout-balanced lattice filters, respectively, under conjugate matching. Notably, all prototypes feature compact footprints smaller than 1.3 mm\textsuperscript{\textbf{2}}. These results highlight the potential of XBAR-based lattice architectures to enable low-loss, wideband acoustic filters for compact, high-performance RF front ends in next-generation wireless communication and sensing systems, while also identifying key challenges and directions for further optimization.
\end{abstract}


\section{Introduction}
\IEEEPARstart{M}{odern} wireless communications and sensing are increasingly data-hungry and pervasive, driving the need for compact systems with high power and spectral efficiency. Among key components, acoustic bandpass filters (BPFs) offer low insertion loss, excellent frequency selectivity, and extreme miniaturization, making them indispensable in modern radio-frequency front ends (RFFEs) operating under multi-standard, spectrum-crowded, and energy-constrained environments \cite{Hagelauer 2023}.

Bulk acoustic wave (BAW) filters are strong candidates for applications beyond 6 GHz, where wavelength-scaling limits the operation of surface acoustic wave (SAW) filters. Although BAW filters, particularly film bulk acoustic resonators (FBARs), are commercially successful, their ultrathin piezoelectric layers, heavy metallization loading, and intrinsic mechanical losses hinder scaling beyond 10 GHz \cite{Vato2026}. In contrast, laterally excited bulk acoustic resonators (XBARs) in suspended thin-film lithium niobate (TFLN) enable frequency scaling beyond 10 GHz, and even above 100 GHz, while maintaining high electromechanical coupling ($k$\textsuperscript{2}) and high quality factor ($Q$) \cite{Lu2025}. Owing to the exceptional piezoelectric and anisotropic properties of TFLN, XBAR-based filters have demonstrated flexible bandwidth design, low loss, and excellent scalability for emerging mobile applications \cite{Gao2021, Barrera2023, Barrera2024-1, Cho2024, Barrera2024-2, Anusorn2025, Barrera2025, Barrera2026}. 

\begin{figure}[tbp]
    \centerline{\includegraphics[width=0.95\linewidth]{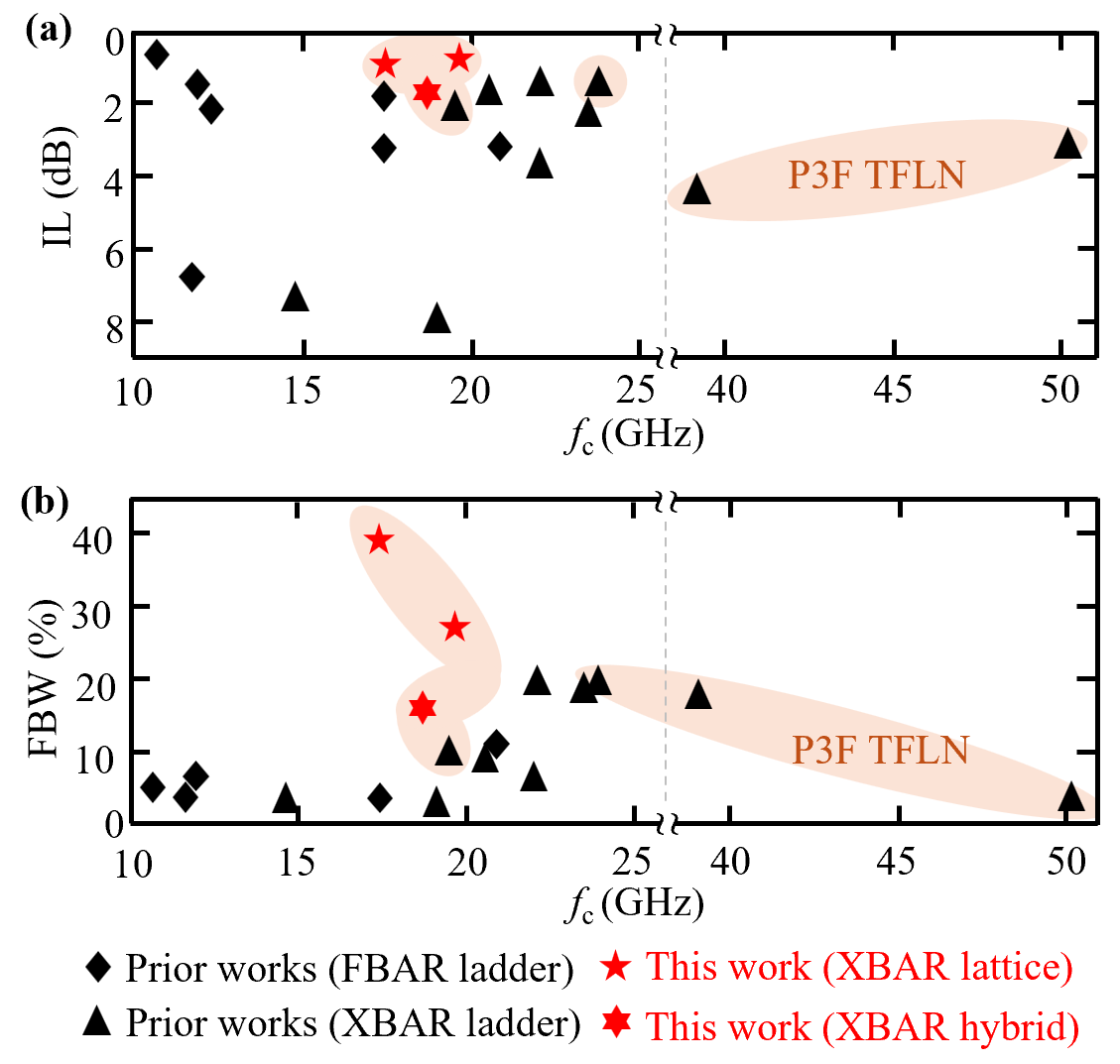}}
    \vspace*{-3mm} 
    \caption{Survey of (a) IL and (b) FBW in acoustic filters above 10 GHz.}
    \vspace*{-5mm} 
    \label{fig1-SoA}
\end{figure}

Periodically poled film (P3F) TFLN, an engineered layered structure, further enhances $k$\textsuperscript{2}, which directly governs the achievable fractional bandwidth (FBW), and improves frequency scalability relative to single-layer films \cite{Kramer2025}. A maximum 3-dB FBW of 19.4\% has been reported for bi-layer P3F TFLN ladder filters \cite{Cho2024}; however, the ladder topology inherently limits further FBW expansion \cite{Heighway1994}. To address this constraint, we present the first lattice XBAR filters to the best of our knowledge. Two lattice filter prototypes exhibit competitive insertion loss (IL) and FBW among state-of-the-art acoustic filters beyond 10 GHz \cite{Gao2021, Barrera2023, Barrera2024-1, Cho2024, Barrera2024-2, Anusorn2025, Barrera2025, Barrera2026, Kochhar2023, Liu2024, Izhar2025, Fiagbenu2025}, as summarized in Fig.~\ref{fig1-SoA}. Furthermore, we built the first practical implementation of a hybrid ladder-lattice XBAR filter.

\section{XBARs in P3F TFLN}
\label{Sec:XBAR}

\begin{figure}[tbp]
    \centerline{\includegraphics[width=0.95\linewidth]{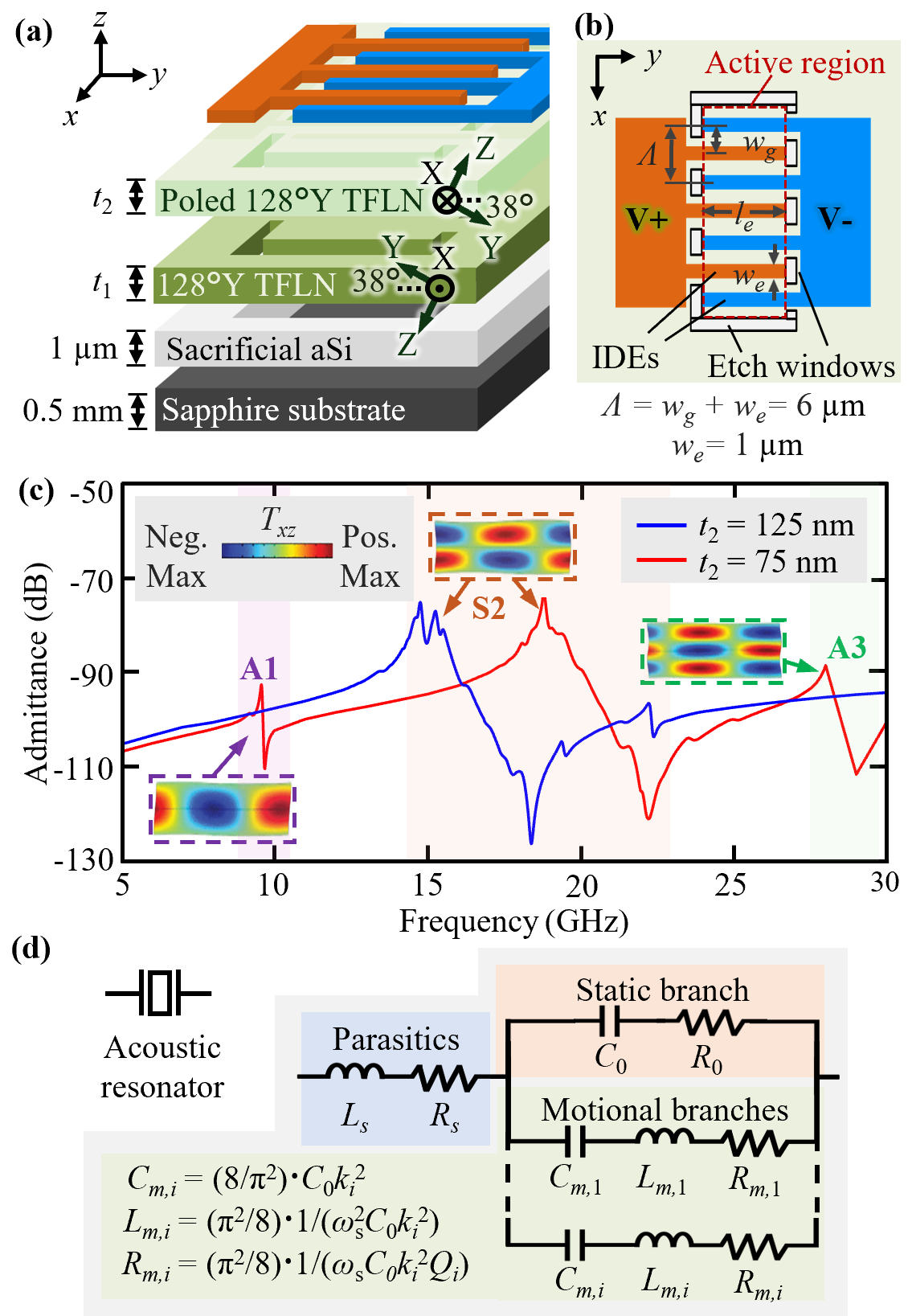}}
    \vspace*{-2mm} 
    \caption{(a) Exploded view and (b) top view of an XBAR implemented in bi-layer P3F 128$^{\circ}$Y-cut TFLN, indicating key dimensional parameters (not to scale). (c) Simulated frequency responses of XBAR unit cells with a fixed bottom TFLN thickness of $t_1 = 110$ nm and varying top-layer thicknesses $t_2$; the corresponding stress mode patterns ($T_{xz}$) are shown adjacent to their associated antiresonances. (d) mBVD model for acoustic resonators exhibiting multiple vibrational modes.} 
    \label{fig2-XBAR}
    \vspace*{-5mm} 
\end{figure}

Following the design methodology in \cite{Anusorn2025}, we first study XBARs in 128$^{\circ}$Y P3F TFLN, as illustrated in Fig.~\ref{fig2-XBAR}(a), using unit-cell finite-element analysis (FEA) in COMSOL Multiphysics. The material axes $X$$-$$Y$$-$$Z$ of each TFLN layer are defined relative to the device coordinates $(x, y, z)$. Notably, the Euler angles for 128$^{\circ}$Y P3F TFLN are $(0, -38^{\circ}, 0)$. The dimensions of the interdigitated electrodes (IDEs), including their electrode width ($w_e$) and length ($l_e$), gap width ($w_g$), and electrode periodicity ($\mathit{\Lambda} = w_e+w_g$), of the XBAR are indicated in Fig.~\ref{fig2-XBAR}(b). 

Fig.~\ref{fig2-XBAR}(c) presents the frequency-domain FEA of the XBAR unit cell with a fixed bottom-layer thickness of $t_1 = 110$ nm and two different top-layer thicknesses, $t_2 =$ 75 nm and 125 nm. Only symmetric Lamb modes (e.g., S2, S6) can be excited in a P3F platform with uniform layer thickness due to polarity constraints \cite{Lu2020}. Nevertheless, thickness mismatches violate these constraints, thereby exciting antisymmetric modes (e.g., A1, A3). This inherent characteristic directly influences the behavior of acoustic filters whose frequency scaling is achieved by trimming only one layer of the P3F stack. To account for multiple vibrational modes, a modified Butterworth-Van Dyke (mBVD) model with multiple motional branches [Fig.~\ref{fig2-XBAR}(d)] is employed for parameter extraction.

\section{Lattice Acoustic Filter Design}

\begin{figure}[tbp]
    \centerline{\includegraphics[width=0.95\linewidth]{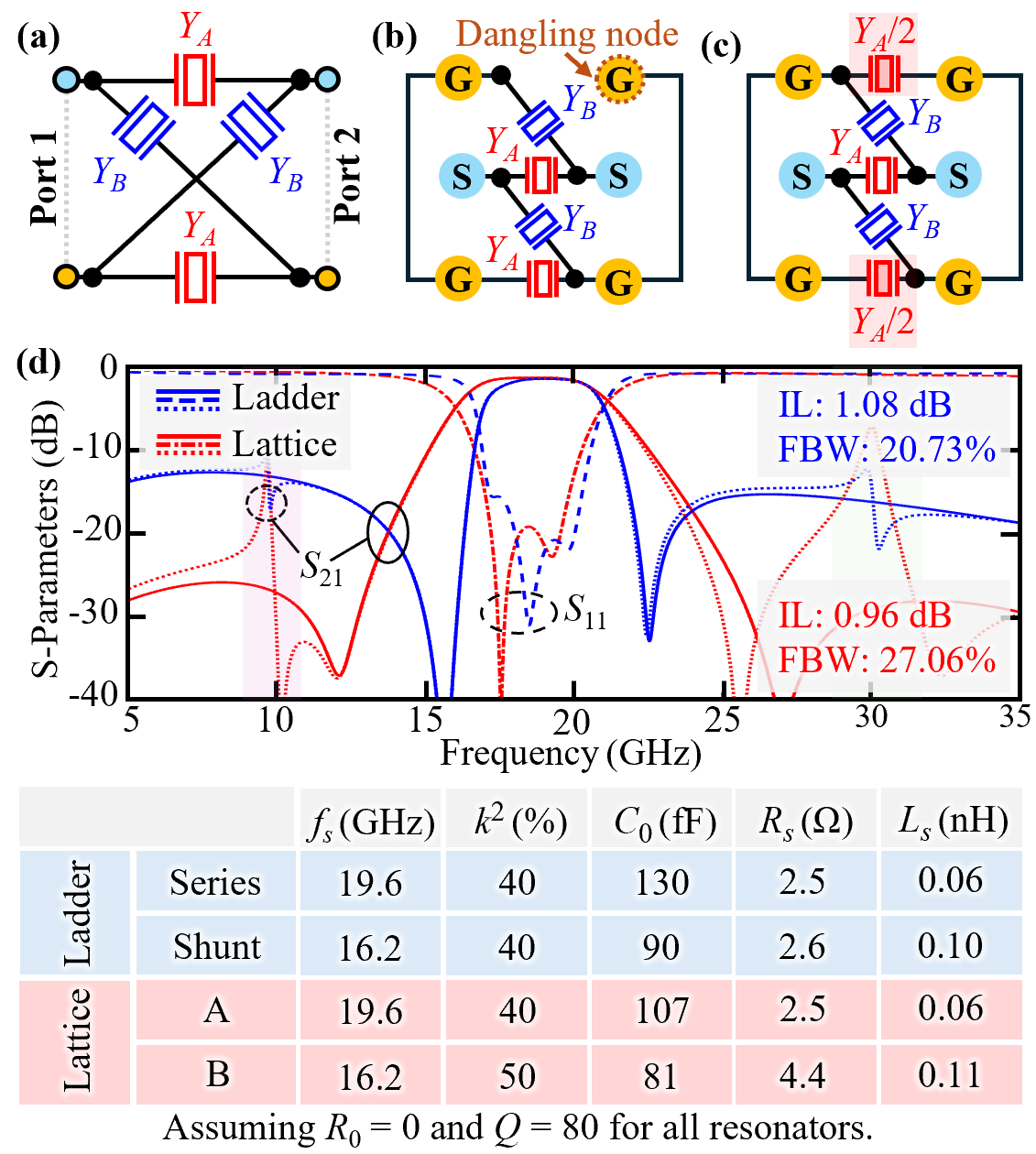}}
    \vspace*{-2mm} 
    \caption{(a) Two-port lattice network. Proposed planar-interconnect implementations: (b) direct lattice and (c) layout-balanced lattice configurations. (d) Simulated 50-$\Omega$ filter responses based on the mBVD circuit model, with a summary of the design parameters. Dotted $S_{21}$ curves illustrate the impact of parasitic A1 and A3 modes on the filter responses.}
    \vspace*{-5.5mm} 
    \label{fig3-FilterDesign}
\end{figure}

A two-port lattice interconnection of acoustic resonators consists of two identical port-connected resonators $Y_A$ and two identical cross-connected resonators $Y_B$ [Fig.~\ref{fig3-FilterDesign}(a)]. The direct lattice design, using planar routing and ground-signal-ground (GSG) probing pads, is shown in Fig.~\ref{fig3-FilterDesign}(b). This layout is inherently asymmetric, resulting in one dangling ground node. To address this, a layout-balanced configuration is proposed, where one port-connected resonator is divided into two equal sections to form complete connections to all nodes [Fig.~\ref{fig3-FilterDesign}(c)]. These configurations are equivalent as long as the admittance between the ground nodes is $Y_A/2$.

The filter circuits are simulated and optimized in Keysight Advanced Design System (ADS) using the extracted mBVD models of the incorporated resonators under balanced operation. To the best of our knowledge, robust analytical synthesis methodologies have only been reported for ladder and transversal acoustic filters \cite{Cano2024, Cano2025, Cano2026}; therefore, formal analytical synthesis for lattice acoustic filters is left for future investigation. Fig.~\ref{fig3-FilterDesign}(d) compares the simulated 50-$\Omega$ responses of the proposed lattice filters with those of a third-order ladder filter employing two shunt resonators as a reference. The corresponding design parameters are summarized below the plot. Notably, layout-induced parasitics and losses, modeled by $R_s$ and $L_s$, are estimated based on prior studies \cite{Barrera2023, Barrera2024-1, Cho2024, Barrera2024-2, Anusorn2025}, and only the target S2 mode is initially considered to limit design complexity. For completeness, however, the filter responses under the presence of additional A1 and A3 modes are also simulated and shown as dotted $S_{21}$ curves. These results clearly indicate that lattice filters are more susceptible to unwanted modal responses. Despite the spurious sidebands, the lattice filter still outperforms the ladder filter in terms of IL, FBW, and overall out-of-band (OoB) rejection.  

\section{Measured Filter Results and Discussions}

Fabrication is perform on a P3F bi-layer TFLN/aSi/sapphire wafer supplied by NGK Corperation, following the process flow described in \cite{Anusorn2025}. To tune the resonant frequencies of certain resonators, 50 nm of the top P3F layer is etched via ion milling \cite{Vato2023}. IDEs and buslines with a thickness of 350 nm are formed by aluminum (Al) evaporation, followed by an additional 550 nm Al thickening on the buslines to reduce series resistance. To suspend the active regions of the XBAR devices, release windows are first defined by ion milling through both TFLN layers, after which the underlying amorphous silicon (aSi) is selectively removed using XeF\textsubscript{2} to release the active region.

\subsection{Direct Lattice Filter}

\begin{figure}[tbp]
    \centerline{\includegraphics[width=0.95\linewidth]{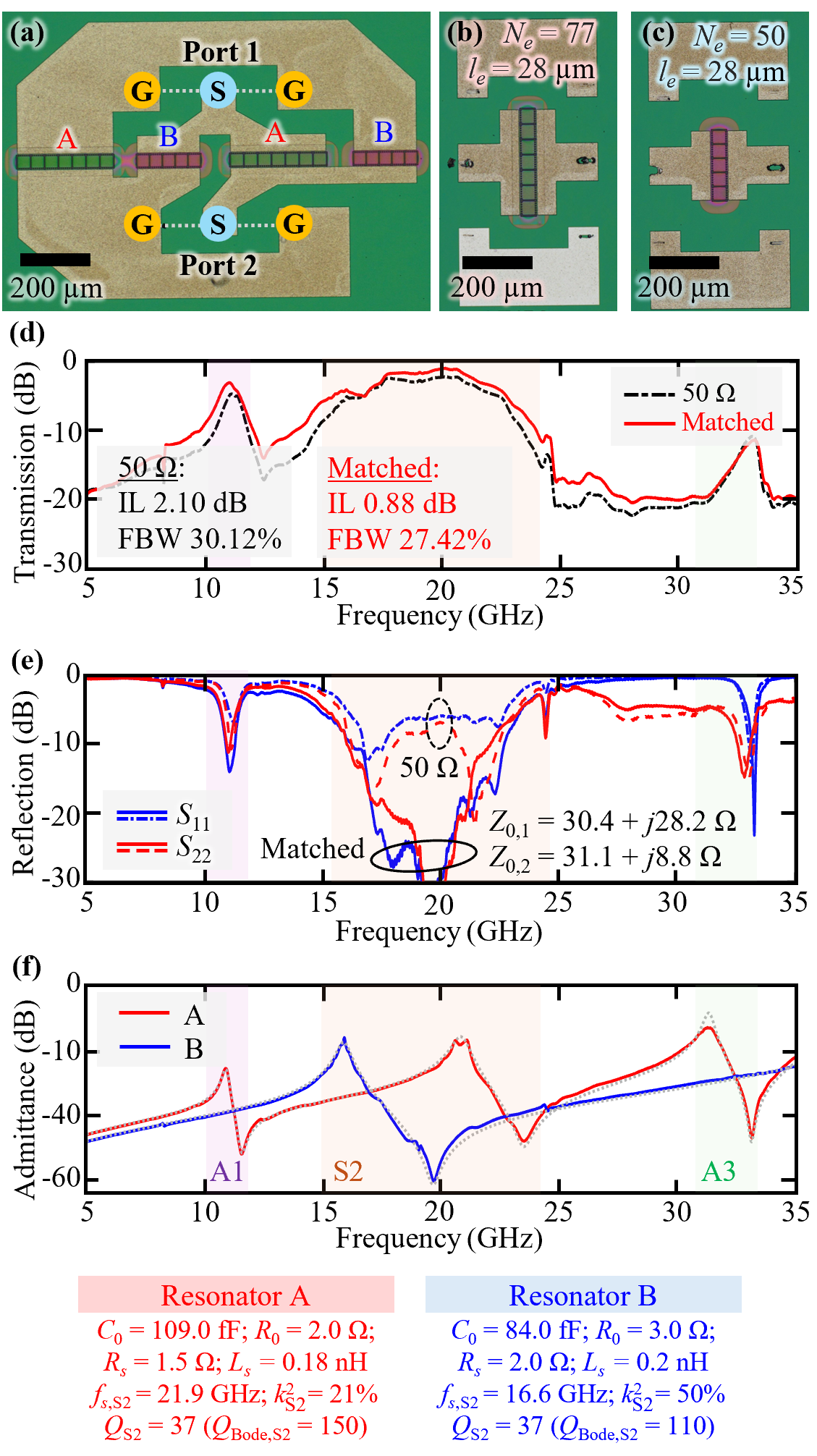}}
    \vspace*{-3mm} 
    \caption{(a) Fabricated direct lattice filter and its standalone resonators: (b) A and (c) B, indicating the number $N_e$ and length $l_e$ of the IDEs. Measured (d) transmission and (e) reflection responses of the prototype filter. (f) Measured admittance of the resonators along with the extracted mBVD model parameters, considering only the S2 mode; the influence of each mode is highlighted.}
    \vspace*{-4mm} 
    \label{fig4-Direct}
\end{figure}

Fig.~\ref{fig4-Direct}(a)$-$(c) shows the fabricated direct lattice filter prototype with a footprint of 1.15 $\times$ 0.81 mm\textsuperscript{2} and its standalone resonators A and B, respectively. The measured transmission and reflection are shown in Fig.~\ref{fig4-Direct}(d) and (e), where both $S_{11}$ and $S_{22}$ are plotted. The measured 50-$\Omega$ responses clearly reveal the asymmetry of this implementation. After applying conjugate matching at each port (30.4 $+$ $j$28.2 $\Omega$ and 31.1 $+$ $j$8.8 $\Omega$ for Ports 1 and 2, respectively), optimal performance is achieved, yielding a minimum IL of 0.88 dB and an FBW of 27\% at a center frequency ($f_c$) of 19.7 GHz. It is worth noting that the optimal complex matching impedance at each port is determined from
\begin{equation}
Z_{0,i} = Z_0 \frac{1 + \Gamma_{m,i}}{1 - \Gamma_{m,i}},
\end{equation}
where the optimal reflection coefficient $\Gamma_{m,i}$ is given by
\begin{equation}
\Gamma_{m,i} = \frac{B_i - \sqrt{B_i^2 - 4|C_i|^2}}{2C_i},
\end{equation}
with
\begin{equation}
B_i = 1 + |S_{ii}|^2 - |S_{jj}|^2 - |\Delta|^2,
\end{equation}
\begin{equation}
C_i = S_{ii} - \Delta S_{jj}^{*},
\end{equation}
and
\begin{equation}
\Delta = S_{11}S_{22} - S_{12}S_{21}.
\end{equation}
Here, $i, j \in \{1,2\}$ with $i \neq j$ \cite{Pozar}.

The measured admittance of each standalone resonator is shown in Fig.~\ref{fig4-Direct}(f), together with the extracted mBVD model parameters. Although the extracted $C_0$ is close to the design value, modeling and parameter estimation inaccuracies lead to minor discrepancies between the simulated and the measured. In addition, the induced A1 and A3 modes in resonator A contribute to undesirable filter sidebands. Notably, the extracted $R_s$ and $R_0$ values may indicate elevated resistive losses, potentially arising from thin buslines due to damage to the thickened Al layer during the lift-off process and from ion-milling-induced damage in the TFLN, respectively. However, these parameters are obtained solely through curve fitting, and dedicated investigations are required to assess how accurately they reflect the underlying physical loss mechanisms.

\subsection{Layout-Balanced Filter}

\begin{figure}[tbp]
    \centerline{\includegraphics[width=0.95\linewidth]{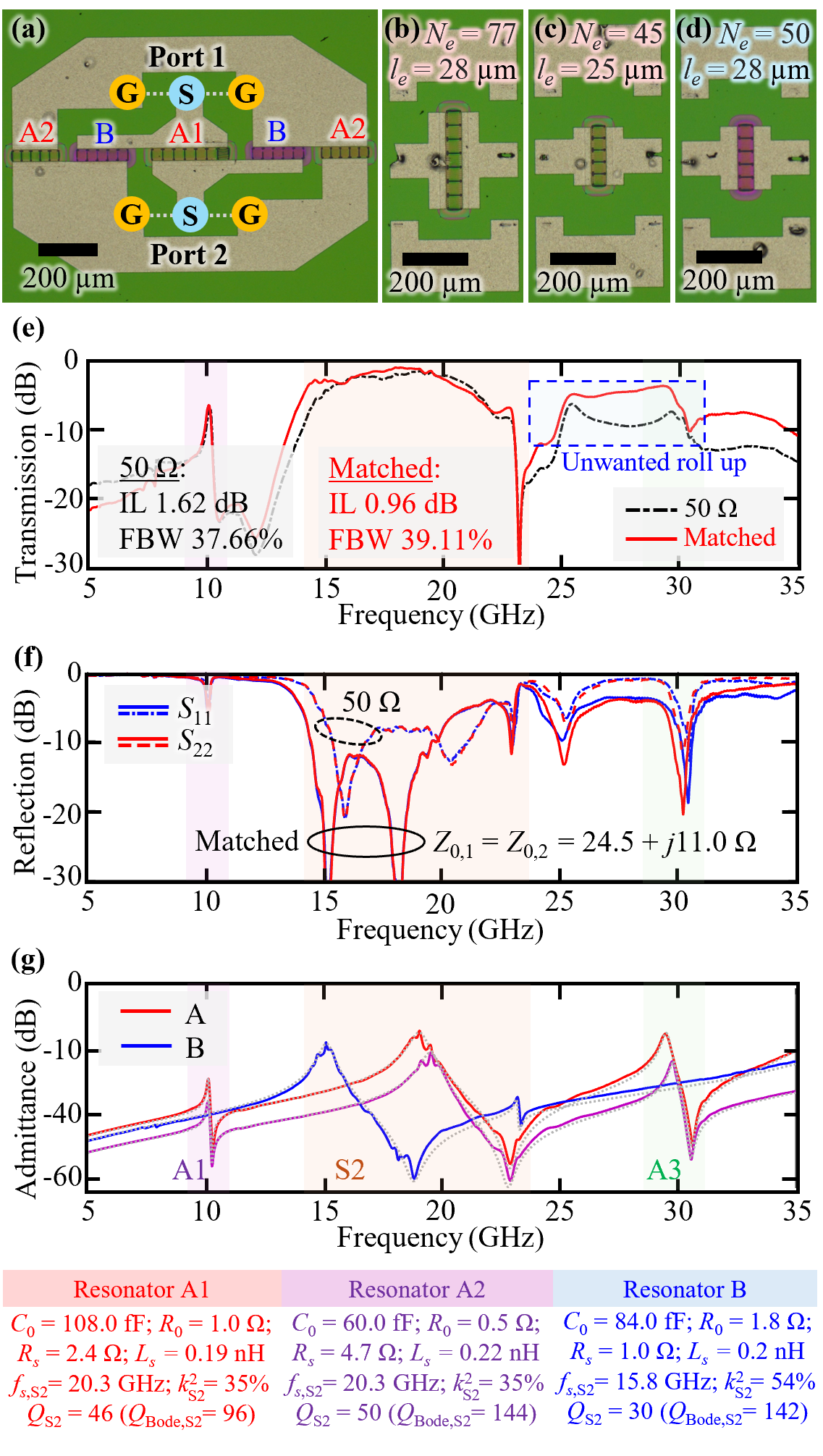}}
    \vspace*{-3mm} 
    \caption{(a) Fabricated layout-balanced lattice filter and its standalone resonators: (b) A1, (c) A2, and (d) B, indicating the number $N_e$ and length $l_e$ of the IDEs. Measured (e) transmission and (f) reflection responses of the prototype filter. (g) Measured admittance of the resonators along with the extracted mBVD model parameters, considering only the S2 mode; the influence of each mode is highlighted.}
    \vspace*{-4mm} 
    \label{fig5-LB}
\end{figure}

The fabricated layout-balanced lattice filter prototype, with a footprint of 1.29 $\times$ 0.85 mm\textsuperscript{2}, along with its constituent resonators A1, A2, and B, is shown in Fig.~\ref{fig5-LB}(a)$-$(d). The measured transmission and reflection are presented in Fig.~\ref{fig5-LB}(e)$-$(f). Notably, the symmetric layout enables a more practical implementation of matching networks. With conjugate matching applied, the filter achieves a minimum IL of 0.96 dB and a wide FBW of 39\%, highlighting the advantages of the lattice topology. A different $f_c$ of 17.7 GHz in this implementation is attributable to local variations in the wafer's LN thickness.

Nevertheless, in addition to unwanted responses induced by the A1 and A3 modes, imperfect resonator splitting (i.e., $Y_{A1} \neq 2Y_{A2}$), as illustrated in Fig.~\ref{fig5-LB}(g), results in an undesirable roll-up in $S_{21}$, as highlighted in Fig.~\ref{fig5-LB}(e). Furthermore, the series-resonance peak locations of XBARs shift with resonator size, even when the series-resonance frequency $f_s$ is identical, due to layout-induced electromagnetic resonances. This effect introduces an additional challenge in the accurate modeling and design of XBAR filters at mmWave \cite{Barrera2024-2, Anusorn2025, Barrera2026}. Techniques and in-depth analyses to mitigate these effects and further optimize the performance of the layout-balanced lattice configuration will be explored in future work.

\section{Conclusion}

This work presents the first practical implementation of lattice XBAR filters. The measurement demonstrates the potential of acoustic filters with low IL and wide FBW at frequencies above 18 GHz. The results also reveal that practical challenges arise from accurate modeling of acoustic devices in the presence of EM resonance, device fabrication, and the synthesis of feasible filters that account for such nonidealities. With continued efforts, the proposed design could yield compact, high-performance RF front-end solutions for next-generation wireless communication and sensing applications.

\section*{Acknowledgment}

The authors thank Dr.\ Ben Griffin, Dr.\ Todd Bauer, Dr.\ Zachary Fishman, Dr. Tzu-Hsuan Hsu, Dr. Harshvardhan Gupta, Dr. Omar Barrera, and Mr. Jack Kramer for their helpful and inspiring discussions.  


\end{document}